\documentclass[final,5p,times,twocolumn]{elsarticle}

\usepackage{float}

\usepackage[all]{nowidow}
\usepackage{csquotes}

\usepackage{mathtools}

\usepackage[detect-all]{siunitx}
\usepackage{dcolumn}
\usepackage{booktabs}
\usepackage{color}

\biboptions{sort&compress}
\usepackage[all]{nowidow}
\usepackage{amsmath}
\usepackage{amssymb}

\journal{Physics Letters B}

\begin{document}

\begin{frontmatter}



\title{Predictions for charged-hadron production in $p$-O collisions at LHC energies}


\author{L.~Konrad$^{1}$, P.~Schulz$^{2}$ and G.~Wolschin$^{1}$}
\noindent
{{
$^1$Institute for Theoretical Physics, Heidelberg University}\\
\hspace{.6cm}        {Philosophenweg 16}\\
      {Heidelberg~
          {69120}\\ 
         {Baden W\"urttemberg},
          {Germany}}\\\\
          \hspace{.6cm} 
          {{$^2$SAP AG}\\
          {Walldorf~
          {69190}\\ 
         {Baden W\"urttemberg},
          {Germany}}\\\\\\\\

\begin{abstract}
We present predictions for centrality-dependent charged-hadron production in $p$-O collisions at top LHC energies  based on a nonequilibrium-statistical relativistic diffusion model. Colour-glass condensate initial conditions are used in a three-sources momentum-space model for gluon-gluon, quark-gluon and gluon-quark sources.
Our results are to be compared with forthcoming Run\,3 pseudorapidity distributions for $\sqrt{s_{NN}}=9.618$ TeV $p$-O collisions.
\end{abstract}



}}

\end{frontmatter}


\section{Introduction}
\label{intro}
Whereas relativistic heavy-ion collisions  {involving} asymmetric systems such as d-Au at the BNL Relativistic Heavy Ion Collider  {(RHIC) or $p$-Pb} at the CERN Large Hadron Collider (LHC) have been investigated extensively in the past \cite{alver11,alice23}, collision systems that are of interest for astrophysics have only recently come to the attention of accelerator physicists.
Particularly intriguing are $p$-O collisions at top LHC energies of $\sqrt{s_{NN}}=9.618$ TeV, which have been investigated experimentally during Run\,3 in July 2025,
with data analysis expected to be completed in early 2026. In this Letter, we present a prediction for centrality-dependent charged-hadron pseudorapidity distributions, with a focus on the production of pions, kaons and protons (and their antiparticles). 

We account for  
hadron production as occuring from three sources in momentum space: a forward-going source arising from the leading particles and the interactions of  {their} partons with  {those} of the backward-going nucleus, a central-rapidity source \cite{Bjorken-1983-Phys.Rev.D27}  {mainly attributable} to gluon-gluon interactions, and a backward-going source caused by the excited fragment participants of the backward-going nucleus. The relative contributions of the three sources change substantially as function of centrality. In particular, the role of the fragmentation sources becomes more pronounced in peripheral collisions, where they can produce an inversion of the maximum of the particle-production amplitude from backward to forward in heavier collision systems such as $p$-Pb, as has been observed experimentally by the ALICE collaboration in $\sqrt{s_{NN}}=5.02$ TeV $p$-Pb collisions at the LHC \cite{alice23}, and interpreted theoretically in \cite{sgw24}.

Here, we connect the relativistic diffusion model  (RDM) with elements from nonperturbative QCD. The initial states in the three sources are represented as color-glass condensate (CGC) states, corresponding to valence-quark -- soft-gluon interactions as accounted for in the fragmentation sources for stopping \cite{mtw09,hgw20,hhw23}, and gluon-gluon interactions for the central source. The subsequent partial thermalization of the fragmentation sources is then calculated from a linear Fokker--Planck equation (FPE), but -- different from our earlier phenomenological calculations that used analytical FPE solutions \cite{gw13,gw15} -- numerical methods must be used for its solution because of the more involved initial conditions. The transport parameters governing the time evolution are estimated from our results in $p$-Pb collisions at  $\sqrt{s_{NN}}=5.02$ and $8.16$ TeV, where they were determined in $\chi^2$-minimizations with respect to the available ATLAS and ALICE  {data \cite{sgw24}.
The central gluon-gluon source produces particles and antiparticles in equal amounts and hence, is not relevant for the stopping process, but generates the largest fraction of charged hadrons in central collisions.  We calculate the yield from the color-glass distribution since partial thermalization is less pronounced as compared to the fragmentation sources, where sizable drift and diffusion is actually observed in $p$-Pb. With this model, we predict the respective role of the forward- and backward fragmentation sources, and the central-rapidity gluon source as functions of centrality in the full pseudorapidity range for $p$-O at the highest available LHC energy.

The CGC initial states for the central-rapidity source based on $k_{T}$-factorization, and  for the two fragmentation sources based on hybrid factorization are given in the next section. The diffusion-model approach to the subsequent time evolution of the fragmentation distribution functions in rapidity space and the numerical solution of the corresponding FPE with the CGC initial conditions is considered in Section\,3. Pseudorapidity distributions for produced particles are discussed in Section\,4,  with results for charged-hadron production for $p$-O collisions at $\sqrt{s_{NN}}=9.618$ TeV for seven centrality classes and minimum bias presented in Section\,5. Conclusions are given in Section\,6.
\section{CGC initial states}
\label{initial}
\subsection{Central source}
For the central gluon-gluon source the initial state is derived using $k_T$ factorization. This approach is commonly used for charged-hadron production from central sources, since it is applicable to high center-of-mass energies $\sqrt{s}$ corresponding to low Bjorken $x$ of the partons. The inclusive cross section was originally presented in \cite{initialgg}, modified for asymmetric $p$-Pb collisions as in \cite{sgw24} by introducing two distinct {additive} subprocesses {that each obey $k_T$ factorization and are proportional to the respective numbers of participants $N_1, N_2$. This model} is now applied to $p$-O collision. The initial state for the central gluon-gluon source with respect to rapidity $y$ and transverse momentum $p_T$ of the produced hadron {becomes}
\begin{align}
    \frac{d^3N_{gg}^h}{dy \, dp_T^2} 
    &= \frac{2\alpha_s}{C_F \, m_T^2} \int_0^{p_T}dk_T^2 \big[ \,
        N_1 \varphi_1(x_1,k_T^2) \varphi_2(x_2, |\mathbf{p}_T-\mathbf{k}_T|^2) \notag \\
    &\hspace{8.4em} + N_2 \varphi_1(x_2,k_T^2) \varphi_2(x_1, |\mathbf{p}_T-\mathbf{k}_T|^2) \big] \text{.}
\end{align}
{The underlying participant scaling dominates the bulk gluon production in soft processes, whereas binary scaling $(\propto N_1*N_2)$ is relevant in hard scatterings such as jets, but plays a minor role for the total gluon yield that determines the bulk of charged-hadron production near midrapidity. Participant scaling reflects the {importance} of low-$x$ gluon fields and gluon saturation in the initial collision stages.}
We calculate the centrality dependence of $N_2$ using a Glauber model for each centrality class. The Casimir factor is $C_F = (N_C^2 - 1) / (2N_C) = \frac{4}{3}$ and the transverse mass {$m_T=(m_p^2+p_T^2)^{1/2}$}. The running of the strong coupling $\alpha_s$ is parametrized \cite{alpha_s} as
\begin{equation}
    \alpha_S(k^2) = \frac{4 \pi}{\beta \ln{(4k^2/\Lambda_\text{QCD}^2 + \mu})}
\end{equation}
with $\beta = 11 - \frac{2}{3} N_f = 9$, for the number of quark flavors $N_f =3$. The QCD-scale parameter is $\Lambda_\text{QCD} =0.241 \,$GeV and $\mu=16.322$. The latter follows from the boundary condition $\alpha_s(\infty) = 0.5$ and regulates the strong coupling at large dipole sizes. 

The distribution depends on the unintegrated gluon distribution functions $\varphi_{1,2}(x, k_T^2)$ for the gluons in the proton $\varphi_1$ and in oxygen $\varphi_2$ with the transverse momentum of the gluon $k_T^2$ and  $x_1 = (m_T/\sqrt{s_{NN}})\,e^y, x_2 = (m_T/\sqrt{s_{NN}})\,e^{-y} $.
{At this stage, the} unintegrated gluon distribution functions do not depend on the impact parameter  $b$. {However, it will later turn out that an additional $b$-dependence  is needed to account for the data, cf. Eq.\,(\ref{eq7}), and the gluon distribution acquires an impact-parameter dependence in our model}. {As in Eq.\,(2) of our previous calculation for $p$-Pb \cite{sgw24}, we} use the Kharzeev-Levin-Nardi model \cite{kln05,Q0Npart}  {for $\phi_{1,2}(x,k_T^2)$ with a prefactor $2C_F/(3\pi^2\alpha_s)$. Whereas the normalization is usually tuned such that theoretical multiplicity distributions match midrapidity data from RHIC and LHC, no such data are available for $p$-O, and we shall normalize the particle-production yield in minimum-bias collisions at midrapidity to Monte-Carlo predictions \cite{MC}.} 
The gluon saturation scale is 
\begin{equation}
    Q_s^2(x) = A^{1/3} Q_0^2 \left( \frac{x_0}{x} \right)^{\lambda}\,.
    \label{saturationscale}
\end{equation}
For the $p$-going source, the parameters are set to $A = 1$ and $x_0 = 1$. According to data from HERA, the parameters for the proton are $\lambda = 0.288$ and $Q_0^2x_0^{\lambda} = 0.097 \,$GeV$^2$. These values were obtained from fits to the experimental results of deep-inelastic electron-proton scattering \cite{Q0scale}. In \cite{saturation}, the consistency of this parametrization with charged-hadron production at LHC energies for Pb-Pb collisions was demonstrated. Therefore, we consider  these values to be applicable in $p$-O collisions as well. 

For the O-going central source, $A=16$ and $x_0 = 1$ is fixed, and the same value $\lambda = 0.288$ is used. Since our previous analysis of the gluon saturation scale in $p$-Pb \cite{sgw24} has shown a centrality dependence of $Q_0^2$, the literature value for $Q_0^2$ from above is not expected to be sufficient for $p$-O collisions as well. Hence, the saturation values determined in \cite{sgw24} for $p$-Pb will be used to estimate suitable values for the present $p$-O calculation.
The gluon distribution functions are modified according to \cite{gdfcorr}
$ \hat{\varphi}(k,x) = (1-x)^4 \varphi(k,x)$
to avoid unphysical contributions from large values of $x$.

\subsection{Fragmentation sources}
\label{sec:fragmentation sources}
The two fragmentation sources for charged hadron production arise from interactions between valence quarks and soft gluons. Due to the asymmetric system, there are two distinct fragmentation sources. For the $p$-going source, interactions between the valence quarks of the proton and the gluons of the oxygen ions are considered, and vice versa for the O-going side. For the O-going side, the valence-quark distribution function $f_{q/A}$ is scaled with the centrality-dependent number of participating nucleons $N_\text{part}$, which is again determined with a Glauber code
\begin{align}
    x_A f_{q/A} = N_\text{part} x_1 f_{q/{{p}}}\, \text{.}
\end{align}
For the fragmentation sources the initial distributions depending on rapidity $y$ and transverse momentum $p_T$ are determined using the method of hybrid factorization. This combines features of $k_T$ factorization and collinear factorization. In this framework, the cross section is determined by treating one incoming parton with small Bjorken $x$ and larger transverse momentum in $k_T$ factorization and the other initial parton with larger Bjorken $x$ and smaller transverse momentum in collinear factorization. The cross section includes the valence-quark distribution function and the gluon distribution for the separately treated incoming nuclei \cite{hybrid}.

The color-glass condensate state for single-inclusive hadron production in asymmetric proton-nucleus scattering is \cite{dhj06,initialfrag}
\begin{align}
    \frac{d^3N_{qg}^h}{dy \, dp_T^2} = \frac{K}{(2 \pi m_T)^2} \int_{x_F}^1 \frac{dz}{{z^2}} D_{h/q}(z,\mu^2_f) x_1 f_{q/p}(x_1,Q_f^2) \varphi(x_2,q_T^2)
\end{align}
with Bjorken $x$ of the valence quark $x_1$ and that of the soft gluon $x_2$. The produced hadrons are $h= \pi$, K, $p$ ,and their antiparticles. $Q_f^2$ is the factorization scale, $f_{q/A}(x_1,Q_f^2)$ is the quark distribution function, and $\varphi(x_2,q_T^2)$ the gluon distribution function.  {Here, we use parton distribution functions 
in the Martin-Stirling-Thorne-Watt  parametrization
(MSTW 2008, \cite{mstw09}).} The Feynman $x_F$ is defined by the transverse momentum of the parton $k_T$ and that of the produced charged hadron $p_T$,
$ x_F = xp_T/k_T$.
The fraction of quark energy carried by the produced hadron is
   $ z(x) = x_F/x$,
with the boundary conditions $z(1)=x_F$ and $z(x_F)=1$, and the associated differential $dz$,
 $   dx/x_F= - (x^2/x_F^2) dz = - dz/z^2$.
The gluon distribution function depends on the effective transverse momentum 
$  q_T = m_T/z=\sqrt{(p_T^2+m^2)}/z$.
The fragmentation function $D_{h/q}(z, \mu_F^2)$ gives the probability that a parton fragments into a hadron carrying the fraction $z$ of the quark's energy{, and we use the updated AKK08 results \cite{akk08}, with parameters provided separately for each hadron species, and symmetrical production of both hadrons and antihadrons}. To obtain the minimum-bias cross section for an effective impact parameter $\langle b \rangle$, the factorization scales of the functions are set to $Q_f^2=p_T^2$ and $\mu_F^2 = p_T^2$. Higher-order corrections for additional dynamical effects can be included in the factor $K$, which is set to $K=1$ here.
{For the forward/backward initial particle production, we take into account only the distribution of incoming quarks on the large-$x$ side, thus neglecting the small contributions of incoming gluons and seaquarks: A similar strategy had previously been successful \cite{mtw09} when calculating the Pb-Pb net-proton (``stopping") distributions in comparison with SPS data.} 

The centrality-dependent numbers of participants are obtained using a Glauber code \cite{GlauberGIT} that is based on \cite{GITGlauberpaper}. 
For the inelastic nucleon-nucleon cross-section, we use $\sigma_{NN} = 73\,$mb for 9.62 TeV $p$-O. This value is determined from LHCb data for the inelastic proton-proton cross-section \cite{ppcrosssec}, in accordance with measurements by  ALICE, ATLAS, and TOTEM. 
We take an exponential profile for the charge density distribution of the proton,
$ f(r) = \exp \left( -r/R \right)$ \cite{GlauberGIT}.
For the oxygen nucleus, we use a three-parameter Eerhart-Fermi distribution \cite{GITGlauberpaper} 
\begin{align}
    f(r) = \frac{1 + W \frac{r^2}{R^2}}{1 + \exp(\frac{r-R}{a})} 
\end{align}
with $R=2.608 \,$fm for the half-density radius, $a=0.513 \,$fm for the skin parameter, and $W=-0.051$ for the suppression near $r=0$. 

The resulting mean numbers of participants $\langle N_\text{part} \rangle = 1 + \langle N_\text{part}^\text{O}\rangle$ are shown in Table \ref{tab1}  for seven centrality classes with the corresponding impact parameters $b$.  In minimum bias (0-100\%) collisions with impact parameters $b=0-4.2$ fm, the average number of participants is $\langle N_\text{part} \rangle=4.2$. Depending on the Glauber codes that the experimental collaborations use when analyzing their data, these values may have to be adapted when comparing model results and forthcoming data.

 Our results for $p$-Pb \cite{sgw24} have shown that a centrality-dependence of the gluon saturation scale is required in order to fit  ATLAS charged-hadron pseudorapidity data at $\sqrt{s_{NN}} = 5.02 \,$TeV, as well as ALICE data at $\sqrt{s_{NN}} = 5.02$ and $8.16\,$TeV. The corresponding dependence of the parameter $Q_0^2$ as obtained from a $\chi^2$-minimization can be expressed as 
\begin{align}
\label{eq7}
    Q_0^2 = c \cdot \log(N_\text{part}) + b
\end{align}
with $c= 0.019 \,$GeV$^2$ and $b=-0.006 \,$GeV$^2$. This indicates a significant decrease of $Q_0^2$ for small values of $N_\text{part}$, especially in the relevant range for $p$-O collisions. {With this modification, the corresponding gluon saturation momentum should be regarded as a model-specific parameter, and care should  be taken when comparing results based on this definition with those obtained using the conventional saturation momentum}. Given the lack of theoretical models for the analytical dependence of $Q_0^2$ on $N_\text{part}$, we use the above result from our $p$-Pb analysis also for $p$-O. Table\,\ref{tab1} includes the ensuing centrality-dependent parameters $Q_0^2 (\langle N_\text{part} \rangle)$ that determine the gluon saturation scale according to 
Eq.\, (\ref{saturationscale}) as $Q_\text{s}^2=Q_0^2\times 16^{1/3}/x^\lambda$  with $\lambda = 0.288$ from HERA.

\begin{center} 
\begin{table}[h] 
\centering
\caption{Calculated mean number of participants, $\langle N_\text{part} \rangle$, for different centrality classes and corresponding impact parameters, $b$, in $p$-O collisions at 9.62 TeV. The centrality-dependent parameters $Q_0^2$ determine the gluon saturation scale.\\}
\begin{scriptsize}
\begin{tabular}{c|cccccccc} 
\hline
cent.[\%] & 0-5 & 5-10 & 10-20 & 20-40 & 40-60 & 60-80 & 80-100 \\ \hline
$b \,$[fm] & 0-0.9 & 0.9-1.3 & 1.3-1.9 & 1.9-2.7 & 2.7-3.3 & 3.3-3.8 & 3.8-4.2  \\
$\langle N_\text{part} \rangle$ & 7.0& 6.2& 5.2& 3.9& 2.8& 2.3 & 2.2\\ 
$Q_0^2$ [GeV$^2$] & 0.031 & 0.028 & 0.025 & 0.020 & 0.014 & 0.010 & 0.009\\
\hline
\end{tabular}
\end{scriptsize}
\label{tab1}
\end{table}
\end{center}
\section{Relativistic Diffusion Model}
The color-glass condensate states describe the stopping distributions {through valence quark-gluon interactions} in the initial stages of the collision, which cause already a sizeable shift of the fragmentation distributions towards midrapidity, as well as a broadening. Subsequently, the partial thermalization of the macroscopic variables begins. In the course of this process, the maxima of the distributions in rapidity space {-- which in principle also include small sea quark components that are not treated explicitly --} are shifted even further away from the beam rapidity and are broadened due to interactions and particle creation. Following the assumptions stated in \cite{sgw24}, particles originating from the fragmentation sources are considered to participate in the partial thermalization process.
This process is described based on a relativistic diffusion model, rigorously derived in \cite{saturation} in a framework for Markovian stochastic processes, and successfully applied to many ion collisions, e.g., in \cite{sgw24} for $p$-Pb or in \cite{gw13,gw15} for Pb-Pb. 

In this approach, the time evolution of a particle distribution is governed by a Fokker--Planck equation \cite{HIC,FWHM,extrapol}. It is formulated here as a one-dimensional model for the distribution function $R(y,t)$ in rapidity space,following integration over transverse momenta
\begin{align}
    \frac{\partial}{\partial t}R(y,t) = - \frac{\partial}{\partial y} [J(y)R(y,t)] + D_y \frac{\partial^2}{\partial y^2}R(y,t)
    \label{FPE}
\end{align}
with the rapidity $y= \ln \left[ {(E+p)/(E-p})\right]$ determined by the particle's energy $E$ and its momentum $p$.
The $p$-going beam rapidity is defined as positive and the O-going direction is negative. 
 \begin{figure}[!h] 
\centering 
\includegraphics[width=\columnwidth]{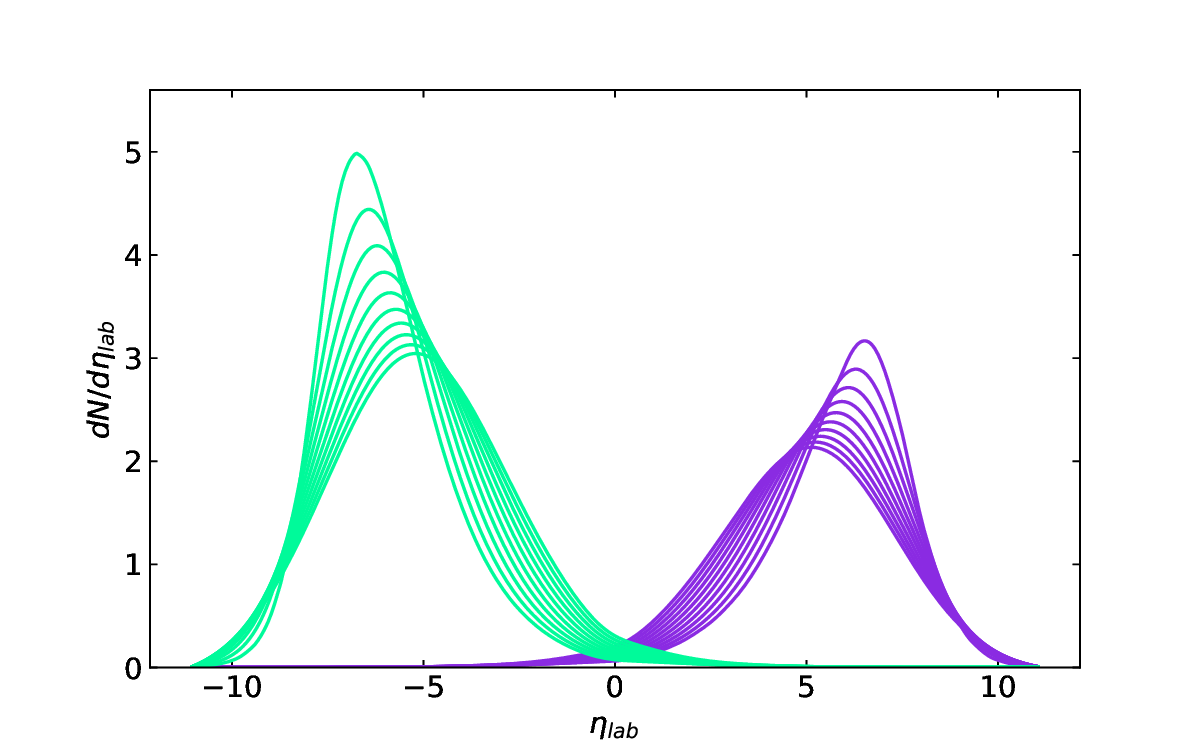} 
\caption{Time evolution of the fragmentation sources in minimum-bias $p$-O collisions at $\sqrt{s_\text{NN}}=9.618$ TeV. The outermost distributions correspond to the color-glass condensate states at the end of stopping, which is taken as the beginning of the diffusion (partial thermalization) process. The innermost distributions with the largest widths are the charged-hadron distributions at the interaction (freezeout) time. Eight equally spaced timesteps are displayed. The distributions in the forward direction (purple) correspond to the $p$-going fragmentation source, and those in the backward direction (green) to the O-going fragmentation source.} 
\label{fig1}
\end{figure}
The rapidity diffusion coefficient $D_y$ ensures the broadening of the distribution that arises from microscopic physics. In our calculations, we assume $D_y$ to be constant over time and rapidity. The drift term $J(y)$ causes the shift of the distribution in rapidity space. To obtain a Maxwell-Jüttner distribution as the stationary solution of Eq.\,(\ref{FPE}), the drift term    $ J(y) = - (m_T D_y/T)\sinh (y)$ would be required \cite{HIC}.
The prefactor depends on the diffusion coefficient $D_y$, the transverse mass $m_T$, and the equilibrium temperature $T$. It results from the corresponding dissipation-fluctuation relation connecting $D_y$ and $J(y)$. The transverse mass is
calculated with the proton mass $m_p$ . 

We use a linear drift term resulting as the leading-order approximation \cite{sgw24}, $J(y) =(y_\text{eq}-y)/\tau_y$,
with the equilibrium rapidity $y_\text{eq}$, and the rapidity relaxation time $\tau_y$ governing the time scale over which the rapidity distribution approaches equilibrium. The Fokker--Planck equation with constant diffusion and linear drift corresponds to the so-called Uhlenbeck-Ornstein process \cite{UO}. The rapidity $y_\text{eq}$ at which equilibrium is reached is zero in symmetric systems, but non-zero in asymmetric systems such as the $p$-O collision. For sufficiently high beam rapidities, given at LHC energies, it can be calculated as  \cite{HIC}
\begin{align}
    y_\text{eq}(b)= \frac{1}{2} \ln \left( \frac{ \langle m_T^{(2)}(b) \rangle}{ \langle m_T^{(1)}(b)\rangle} \right)
\end{align}
with the centrality-dependent average transverse mass $\langle m_T^{1,2}(b) \rangle$.
In equilibrium, the charged-hadron distribution approaches a Maxwell-Jüttner distribution of the form
\begin{align}
    E \frac{dN}{dp^3} \sim E \exp(-E/T) = m_T \, \cosh(y) \exp(-m_T \cosh(y)/T) \text{.}
\end{align}
The particle production in the longitudinal fragmentation sources is obtained by integrating over the transverse mass \cite{sgw24}. For the integration, {cylindrical} symmetry in phase space for transverse momentum and transverse mass is assumed, to obtain 
\begin{align}
    E \frac{dN}{dy}(y,t) = c \int_m^\infty m_T^2 \cosh(y) R(y,t) \, dm_T
\end{align}
where $c$ is a constant that absorbs all other constants appearing in the expression, which are not relevant for the integration. 

The Fokker--Planck equation with linear drift and constant diffusion can be solved analytically for $\delta$- or Gaussian initial conditions. However, this is not possible for the more sophisticated color-glass condensate initial states used here and in \cite{sgw24}. Therefore, a numerical solution of the differential equation is required. The numerical integration is carried out for the two fragmentation sources using a  C++ code based on the finite-element method, originally written for the \mbox{$p$-Pb} system \cite{sgw24} and modified for this prediction. We obtain the distribution functions for the two fragmentation sources, $R_{1,2}(y,t)$, and use the initial state for the central gluon-gluon source, $R_{gg}(y,t)$. Due to the linearity of the Fokker--Planck equation, we can construct the distribution of charged-hadron production as an incoherent superposition of the three sources \cite{sgw24}
\begin{align}
    \frac{dN_\text{ch}}{dy}(\tau_\text{int}) = N_\text{ch}^1 R_1(y, \tau_\text{int}) + N_\text{ch}^2 R_2(y, \tau_\text{int}) + N_\text{ch}^{gg} R_{gg}(y, \tau_\text{int})\,.
    \label{superposition}
\end{align}
The fragmentation sources are integrated up to the interaction (freezeout) time $\tau_\text{int}$. The number of charged hadrons in the fragmentation sources are  $N_\text{ch}^{1,2}$, while $N_\text{ch}^{gg}$ corresponds to the production of charged hadrons in the central source. The amount of produced charged hadrons and the contribution of the sources are discussed in Section \ref{sec:results}. The time evolution of the fragmentation sources in minimum-bias $p$-O collisions at $\sqrt{s_\text{NN}}=9.618$ TeV is displayed in Fig.\,\ref{fig1}.

\begin{figure}[!h] 
\center
	\includegraphics[width=\columnwidth]{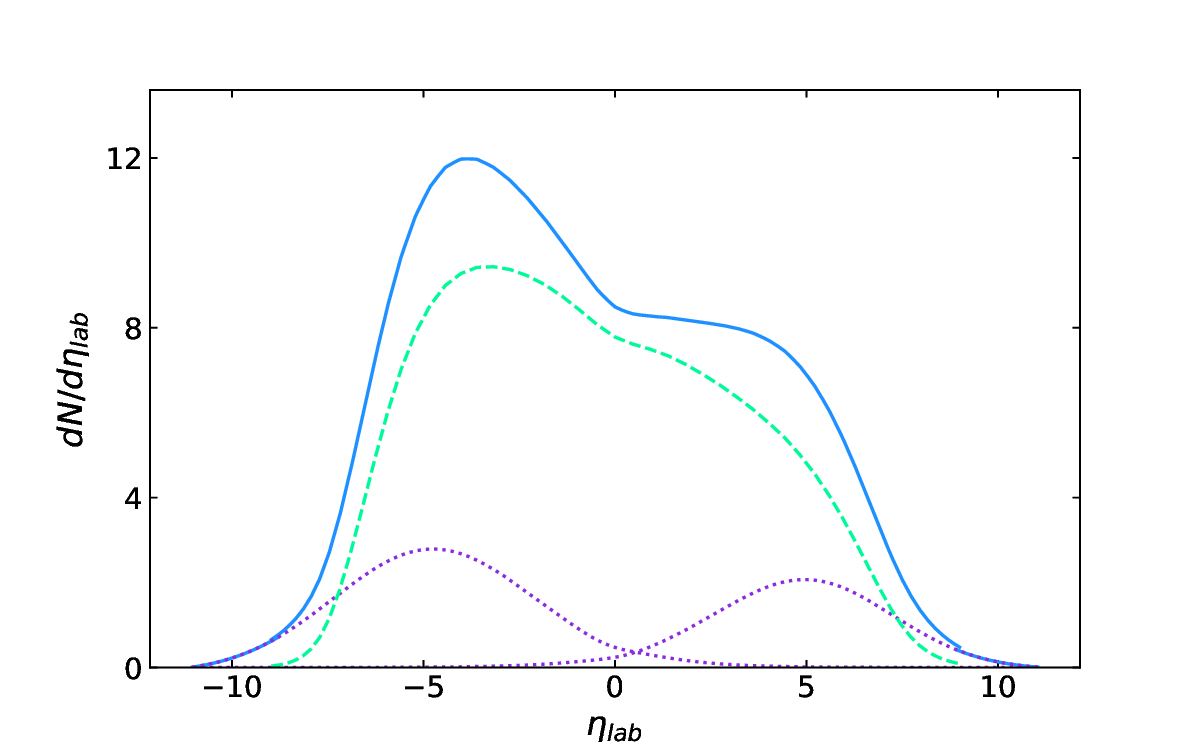}%
	\caption{\label{fig2}%
Calculated minimum-bias pseudorapidity distributions of produced charged hadrons (solid curve) in $p$-O collisions at $\sqrt{s_{NN}}=9.618$ TeV. The dashed curve shows charged hadrons (pions, kaons, protons, and their antiparticles) from the central gluon-gluon source, the dotted curves are the ones from the respective forward- and backward-going fragmentation sources.} 
\end{figure}

To account for the subsequent partial thermalization process, the transport parameters $D_y$, $J(y)$ (which includes $\tau_y$), and the interaction time $\tau_\text{int}$ are required. The absolute timescales $\tau_y$ and $\tau_\text{int}$ are, however, not observables. Instead, the observables are the full widths at half maximum, which are related to $D\tau_y$ and $\tau_\text{int}/\tau_y$, as well as the shifts of the maxima towards equilibrium, which depend on $\tau_\text{int}/\tau_y$. 

Especially for asymmetric systems, the superposition of the sources is very sensitive to the values of the transport parameters, {which depend on the system size $(\propto A^{1/3})$ and the available energy. The latter is rather similar in $p$-Pb $(\sqrt{s_{NN}}$=8.16 TeV) and $p$-O $(\sqrt{s_{NN}}$=9.62 TeV).} A QCD-based calculation in the nonperturbative regime to determine these parameters is{, however, }not feasible. An extrapolation, as done for symmetric systems in \cite{extrapol}, is impossible since no data are available for transport parameters in asymmetric heavy-ion collisions except for $p$-Pb collisions \cite{sgw24} and for d-Au with analytical solutions of the Fokker--Planck equation \cite{biya05}. To estimate the transport parameters for $p$-O collision at $\sqrt{s_{NN}} = 9.618 \,${TeV}, their values in $p$-Pb at $\sqrt{s_{NN}} = 8.16 \,${TeV}, resulting from $\chi^2$-fits to ALICE data, are taken from \cite{sgw24}, and scaled in each centrality class by a factor $\log({A_\text{O}^{1/3})/\log(A_\text{Pb}^{1/3})}\simeq 0.519$, to account for the different system sizes and time scales {consistent with previous investigations of particle production in d-Au, Au-Au, $p$-Pb and Pb-Pb systems \cite{FWHM,extrapol,biya05}}. The resulting parameters are $D\tau_y = 12.5$ and $\tau_\text{int}/\tau_y = 0.26$ for oxygen, as well as $\tau_\text{int}/\tau_y = 0.21$ for the proton. The centrality-dependent transport parameters $D^p\tau_y$ for the proton are shown in Table \ref{tab2}.


\section{Pseudorapidity Distributions}
To obtain a prediction that can later be compared to data, we need to consider the rapidity shift from the nucleon-nucleon center-of-mass frame to the laboratory frame. In the ultrarelativistic limit that is applicable at TeV energies, it is given by
\begin{align}
   \Delta y=\frac{1}{2}\ln\left[\frac{A_1 Z_2}{A_2 Z_1}\right] \,,
    \label{shift}
\end{align}
yielding $\Delta y=0.465$ for $p$-Pb, and $\Delta y=0.347$ for $p$-O, irrespective of energy.
The beam rapidities are given by $|y_\text{beam}^p| = 9.582$ for the proton and $|y_\text{beam}^\text{O}| = 8.889$ for oxygen. The beam rapidity in the nucleon-nucleon frame of reference can be obtained by using $|\mathbf{p}_{NN}| =\sqrt{s_{NN}}/2$ as momentum per nucleon of the colliding beams, resulting in $|y_\text{beam}| = 9.235$. 

By integrating the Fokker--Planck equation for the color-glass initial conditions within the relativistic diffusion model, charged-hadron distributions in rapidity space are obtained for the fragmentation sources. For the central gluon-gluon source, the color-glass initial states are sufficient to account for the charged-hadron distributions in rapidity space. Since data for charged-hadron production at the LHC will be available in pseudorapidity space, a Jacobian transformation is applied to the solutions of the relativistic diffusion model, as well as to the central gluon-gluon source  as detailed in \cite{sgw24}. 
The shape and position  of each source in pseudorapidity is then fixed. However, they still must be scaled to obtain the total yield of charged hadrons and the contribution of the three sources to the total distribution in each centrality class, as indicated in 
Eq.\,(\ref{superposition}). 

To account for the correct contribution of each source, two parameters are considered \cite{sgw24}: the ratio of hadrons produced in the central source to the those produced in the fragmentation sources, $R_{qg}^{gg} = N_{gg} / N_{qg}$, and the ratio of particles originating from the O-going fragmentation source to those from the $p$-going source, $R_{p}^{O} = N_{qg}^\text{O} / N_{qg}^p$.

The latter ratio is obtained by comparing the integrals of the two fragmentation sources over the full pseudorapidity range. The ratio is already fixed by the initial conditions, since the integrals of the fragmentation sources do not change throughout the partial thermalization process.

The contribution of the central source in $p$-O collisions is expected to be {relatively} more significant than that in $p$-Pb.
{In particular, the scaling behavior of the ratio $R_{qg}^{gg}$  can be understood in terms of cold nuclear matter effects, which reduce the relative contribution of the central $gg$ source as compared to the $qg$ and $gq$ fragmentation sources. Such effects depend on the size of the transverse overlap. To quantify the influence of the transverse area, we use a scaling factor proportional to $A^{-2/3}$. Since the maximum particle production amplitude of the minimum-bias distribution cannot be determined a priori in our model, we have normalized it to match the maximum of the corresponding Monte Carlo particle distribution in the midrapidity region \cite{MC}.}
 The contribution of the central source in all other centrality classes was determined by assuming the same decreasing behavior as observed in $p$–Pb collisions at $\sqrt{s_{NN}} = 8.16\,$TeV, with parameter values given in Table \ref{tab2}. 
\begin{figure}[!h] 
	\includegraphics[width=\columnwidth]{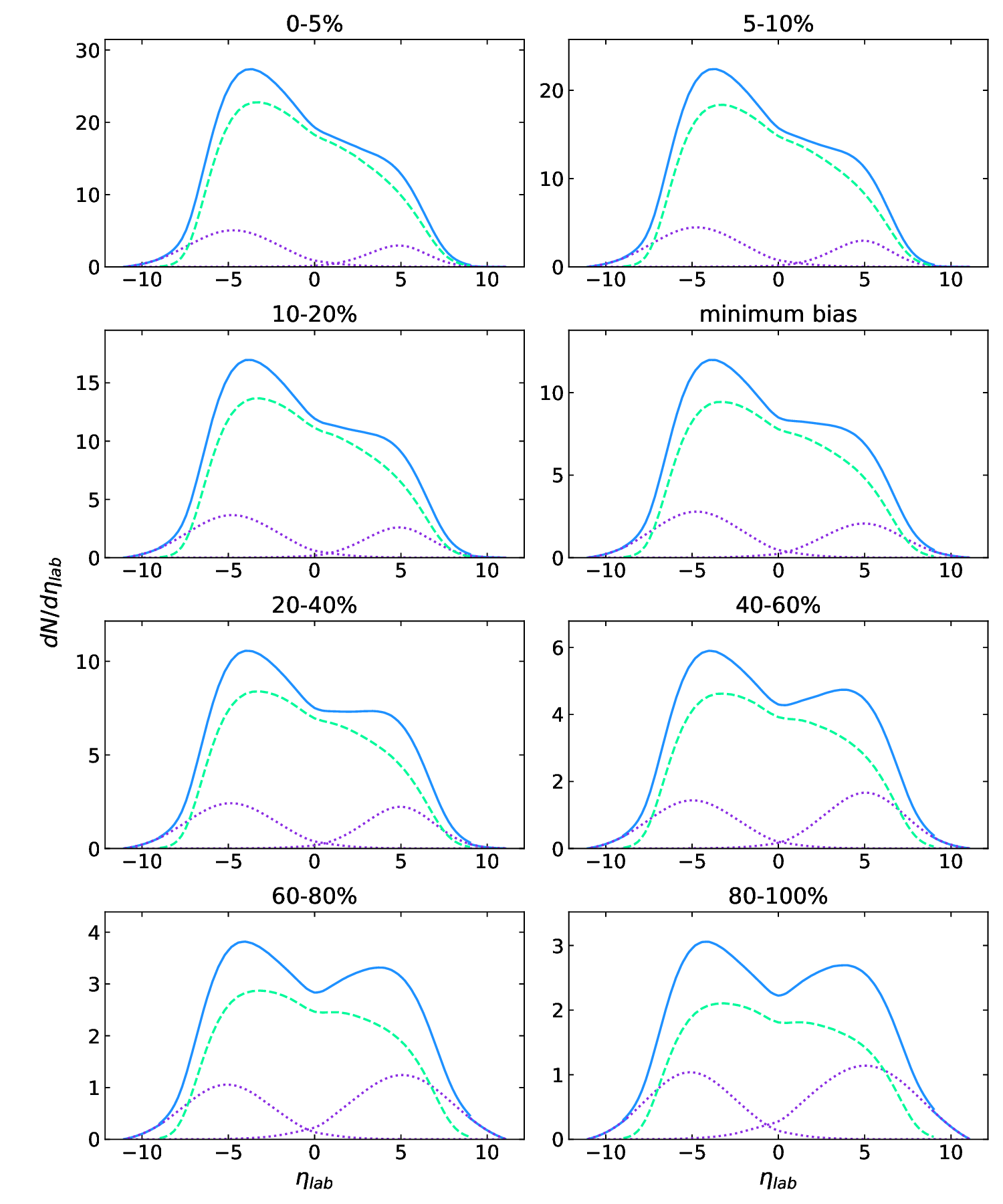}%
	\caption{\label{fig3}%
Calculated pseudorapidity distributions of produced charged hadrons, solid curves) in $p$-O collisions at $\sqrt{s_{NN}}=9.618$ TeV for seven centrality classes and minimum bias (0-100\%). The dashed curves show charged hadrons (pions, kaons, protons, and their antiparticles) from the central gluon-gluon source, the dotted curves are the ones from the respective forward- and backward-going fragmentation sources.}
\end{figure}
\section{Results}
\label{sec:results}
Following the transformation to pseudorapidity space, and scaling the distributions to obtain the expected particle yield, as well as adding the three sources incoherently, the final charged-hadron distributions, which can be compared to data, are obtained. 
The parameters that do not depend on the impact parameter of the collision are for the $p$-going source the gluon saturation scale, $Q_0^2 =0.097 \,$GeV, and the transport parameter, $\tau^{p}_\text{int}/\tau_y= 0.21$. For the source in O-going direction the centrality-independent transport parameters are $D \tau_y =12.5$ and $\tau_\text{int}/\tau_y = 0.26$. An overview of the centrality-dependent parameters is provided in Table \ref{tab2}.
\begin{center} 
\begin{table}[h] 
\centering
\caption{Centrality-dependent parameters used to calculate charged-hadron production in $p$-O collisions: number of participating nucleons, $\langle N_\text{part} \rangle$, gluon saturation scale for the O-going source, $Q_0^2$, and the transport parameter for the $p$-going source, $D^p \tau_y$, ratio of hadrons produced in the central source to those produced in the fragmentation sources, $R_{qg}^{gg}$, and ratio of particles originating from the O-going fragmentation source to those from the $p$-going source, $R_{p}^\text{O}$. $N_\text{ch}$ is  the total number of charged hadrons  produced by all sources in $p$-O collisions at each centrality.\\}
\begin{scriptsize}
\begin{tabular}{c|cccccccc} 
\hline
centrality [\%] & 0-5 & 5-10 & 10-20 & 20-40 & 40-60 & 60-80 & 80-100 \\ 
 \hline
$\langle N_\text{part} \rangle$ & 7.0& 6.2& 5.2& 3.9& 2.8& 2.3 & 2.2\\
$Q_0^2$ [GeV$^2$] & 0.031 & 0.028 & 0.025 & 0.020 & 0.014 & 0.010 & 0.009  \\ 
$D^p \tau_y$ & 3.6 & 3.6 & 5.7 & 6.8 & 10.4 & 18.7 & 24.9 \\ 
$R_{qg}^{gg}$ & 1.98& 1.76& 1.54& 1.32& 1.10 & 0.88& 0.66 \\
$R_{p}^\text{O}$ & 2.47 & 2.18 & 1.81 & 1.33 & 0.94 & 0.78 & 0.77  \\
$N_\text{ch}$ & 264 & 220 & 172 & 115 & 71 & 49 & 40 \\  
\hline
 \end{tabular}
\end{scriptsize}
\label{tab2}
\end{table}
\end{center}
The resulting charged-hadron distributions in pseudorapidity space for minimum-bias $p$-O collisions are shown in Fig.\,\ref{fig2}. In  Fig.\,\ref{fig3}, they are presented for seven centrality classes as given in Table~\ref{tab2}, together with the minimum-bias result. Positive pseudorapidities $\eta$ correspond to the forward $p$-going direction, while negative pseudorapidities correspond to the backward O-going direction. 
The total number of produced charged hadrons decreases from central to peripheral collisions, with the most central collision (0-5\%) yielding about 6.4 times more charged hadrons than in the most peripheral one (80-100\%), as can be seen in Table \ref{tab2}, last line. The values were determined by integrating the charged-hadron distributions over the full pseudorapidity range.

In $p$-Pb collisions, the amplitude of the charged-hadron production in the $p$-going direction becomes larger than the amplitude in the Pb-going direction in very peripheral collisions. The central gluon-gluon source cannot compensate for this effect, although it shows a slight preference towards the Pb-going side even in peripheral collisions, since its contribution in peripheral collisions is too small. This effect has been calculated within the relativistic diffusion model and has been confirmed with LHC data \cite{sgw24}. 
This inversion does not appear in our model calculation for $p$-O collisions. The gluon field of the oxygen nucleus is much weaker than that of lead. Hence, the difference between the gluon fields in the proton and the nucleus is not as large in $p$-O collisions, compared to $p$-Pb. Thus, the fragmentation source in $p$-going direction does not increase as strongly and is compensated by the central source, which is biased towards the O-going direction. 
\section{Conclusions}
In this Letter, we have predicted centrality-dependent charged-hadron production in $p$-O collision at $\sqrt{s_{NN}}=9.618 \,$TeV. The corresponding experiments have been carried out at the Large Hadron Collider in 2025, results will be available for comparison in 2026. We use the relativistic diffusion model as designed for asymmetric collisions. It accounts for charged-hadron production with three sources in momentum space: one central gluon-gluon source and two fragmentation sources. For each source, a QCD-inspired color-glass condensate state as initial state is considered. 

The time evolution of the partial thermalization process for the fragmentation sources has been determined using a Fokker--Planck equation with a linear drift and a constant diffusion parameter. The transport parameters, namely the diffusion parameter and the interaction time, are estimated from values for $p$-Pb collisions at $\sqrt{s_{NN}}=8.16 \,$TeV. Due to the linearity of the Fokker--Planck equation, particle production in the three sources can be added incoherently to obtain charged-hadron distributions in rapidity space. These are transformed to pseudorapidity space using suitable Jacobians and scaled with parameters that control the contributions of each source. We have thus obtained charged-hadron distributions in pseudorapidity space in seven centrality classes as well as the minimum bias distribution. For the smaller $p$-O system, we do not observe an inversion of the particle-production amplitude when going from central to very peripheral collisions.

{When} experimental results for $p$-O will be available, the transport parameters and the centrality-dependent gluon-saturation scale can be adapted to the data {through $\chi^2$ minimizations as in our previous work for $p$-Pb. However, should the model fail to describe the data accurately even with updated parameters, it will be important to identify which of its aspects carry the largest uncertainties, and to reconsider the basic assumptions.
Once the model and its possible updates are} consistent with the new LHC data, it can be used for further analyses and predictions of various heavy-ion collisions. {In any case, the availability of the forthcoming $p$-O} data will broaden our understanding of ultrarelativistic {heavy-ion collisions as well as} cosmic-ray interactions and help refine theoretical models describing them.

\section*{Acknowledgement}
GW is grateful to Jean-Paul Blaizot for a discussion during his stay at ITP Heidelberg.
\bibliographystyle{elsarticle-num}

\end{document}